\documentstyle[12pt]{article}%
\renewcommand{\title}[1]{\large\bf #1\bigskip\medskip\\} 
\renewcommand{\author}[1]{\large #1\\ \smallskip}
\newcommand{\address}[1]{{\normalsize\it #1\\}\bigskip}

\newcommand{\smat}[1]{\mbox{\small $\pmatrix{#1}$}}
\newcommand{\be}{\begin{eqnarray}}
\newcommand{\ee}{\end{eqnarray}} 
\newcommand{\hs}[1]{\hspace*{#1cm}}

\newcommand{\no}{\nonumber}

\def\R{\tilde{R}}
\def\D{\cal{D}}
\def\a{\tilde{a}}
\def\b{\tilde{b}}
\def\P{{\cal P}}
\def\A{{\cal A}}

\def\N{\cal{L}}
\def\i{\mbox{i}}
\def\al{\mbox{$\alpha$}}
\def\U{\mbox{$\cal  A$}}
\def\e{\mbox{$\epsilon$}}
\def\cb{\mbox{$\cal T$}}
\def\la{\mbox{${\ell}\!a$}}
\def\Le{\mbox{${\ell}\!e$}}
\def\lA{\mbox{${\ell}\!A$}}
\def\scr{\scriptsize}
\def\half{\mbox{$\textstyle {1 \over 2}$}}
\def\3half{\mbox{\small$3\over 2$}}

\def\h{\hs{0.5}}
\def\({\biggl(}
\def\){\biggr)}
\def\[{\mbox{$\biggl\lfloor$}}
\def\]{\mbox{$\biggr\rfloor$}}
\def\and{\mbox{\h and \h}}
\def\disp{\displaystyle}

\def\sqbox#1#2#3#4{\setlength{\unitlength}{0.0105in}
  \begin{picture}(20,20)(-#1,-#2)
   \put(0,0){\line(1,0){20}}
   \put(0,0){\line(0,1){20}}
   \put(20,20){\line(-1,0){20}}
   \put(20,20){\line(0,-1){20}}
   \put(7,7){\scr $#3$}
   \put(21,21){\scr $#4$}
\end{picture}}
\def\sqqbox#1#2#3#4#5{\setlength{\unitlength}{0.0105in}
  \begin{picture}(20,40)(-#1,-#2)
   \put(0,0){\line(0,1){40}}
   \put(20,0){\line(0,1){40}}
   \put(0,0){\line(1,0){20}}
   \put(0,20){\line(1,0){20}}   
   \put(0,40){\line(1,0){20}}
   \put(7,27){\scr $#3$}
   \put(7,7){\scr $#4$}
   \put(21,41){\scr $#5$}
\end{picture}}
\def\sqbbox#1#2#3#4{\setlength{\unitlength}{0.0124in}
  \begin{picture}(20,20)(-#1,-#2)
   \put(0,0){\line(1,0){20}}
   \put(0,0){\line(0,1){20}}
   \put(20,20){\line(-1,0){20}}
   \put(20,20){\line(0,-1){20}}
   \put(8,8){\tiny $#3$}
   \put(21,21){\scr $#4$}
\end{picture}}
\def\sqboxd#1#2#3{\setlength{\unitlength}{0.0105in}
  \begin{picture}(40,20)(-#1,-#2)
   \put(0,0){\line(0,1){20}}
   \put(40,0){\line(0,1){20}}
   \put(0,0){\line(1,0){40}}
   \put(0,20){\line(1,0){40}}
   \put(10,7){$\cdots$}
   \put(17,-20){\scr $#3$}
\end{picture}}

\def\Atwon#1#2#3#4{\setlength{\unitlength}{0.0065in}%
\begin{picture}(#1,45)(0,0)
\multiput(0,0)(#1,0){2}{\line(0,1){30}}
\multiput(0,0)(0,30){2}{\line(1,0){#1}}
\put(#1,6){$\hs{0.3}#2$}\put(#3,-22){\scr$#4$}
\end{picture}}

\begin{document}
\begin{flushright} 
ANU Preprint MRR 058-96 
\end{flushright}
\begin{center}
\title{Spin excitations in the integrable open \\quantum group invariant 
    supersymmetric $t$--$J$ model} 
\author{Y.-K. Zhou and M.~T. Batchelor}
\address{Department of Mathematics, School of Mathematical Sciences,\\
         Australian National University, Canberra ACT 0200, Australia}
\begin{abstract} 
The integrable quantum group $spl_q(2,1)$-invariant supersymmetric 
$t$--$J$ model
with open boundaries is studied via an analytic treatment of the Bethe 
equations. An $su(2)$ feature is seen to hold for states 
at or close to half-filling. For these states the 
eigenvalues of the transfer matrix of the $t$--$J$ model satisfy a 
set of $su(2)$ functional relations. 
The finite-size corrections to the relevant eigenvalues, and thus the
surface effect on the spin excitations,  have been 
calculated analytically by solving the functional relations.

\end{abstract}
\end{center}

\subsection{Introduction}\setcounter{equation}{0}

The integrable supersymmetric $t$--$J$ model of strongly correlated 
electrons has a long and interesting history 
(see, e.g., \cite{KE:94,FoKa:93,Ruiz94,E:96,JuKl:96} and refs therein).
More recently, the construction of the integrable version with open boundary 
conditions \cite{FoKa:93,Ruiz94} 
had to await the systematic development of 
boundary integrability \cite{Cherednik,Sklyanin}.
For open boundaries, the integrable Hamiltonian reads 
\be
H&=&-\P\left\{\sum_{j=1}^{L-1}\sum_{\sigma}\left(c_{j,\sigma}^{\dagger}
c_{j+1,\sigma}+c_{j,\sigma}c_{j+1,\sigma}^{\dagger}\right)\right\}\P
\no\\
&-&2\sum_{j=1}^{L-1}\left(S_{j}^{x}S_{j+1}^{x}+S_{j}^{y}S_{j+1}^{y}+
\cos\gamma\;\left(S_{j}^{z}S_{j+1}^{z}-\frac{n_{j}n_{j+1}}{4}\right)\right)
-\cos\gamma\sum_{j=1}^{L}n_{j}\no\\
&+&{\i}\sin\gamma\;(n_{1}-n_{L})-{\i}\sin\gamma\sum_{j=1}^{L-1}(n_{j}S^{z}_{j+1}
 -S^{z}_{j}n_{j+1}) +H_s^+ +H_s^- , \label{H}
\ee
where the boundary fields $H_s^{\pm}$ are dependent on two
arbitrary parameters $\xi_{\pm}$ \cite{Ruiz94}.
The operators $c_{j\pm}$ ($c^{(\dagger)}_{j\pm}$) are spin up or
down annihilation (creation) operators. The 
${\boldmath S}_{j}=(S_j^x, S_j^y, S_j^z)$ are
spin operators and $n_{j}$ the occupation number of
electrons at site $j$. The operator
$\P=\prod_{j=1}^{L}(1-n_{j\uparrow}n_{j\downarrow})$ forbids the double
occupancy of electrons at one lattice site.

We begin by recalling the essential ingredients underlying the integrability 
of the Hamiltonian (\ref{H}).
The $R$-matrix (see, e.g., \cite{FoKa:93,Ruiz94}) is given by
\be
R(v)=\smat{a(v)&0&0&0&0&0&0&0&0\cr
           0&b(v)&0&c_{-}(v)&0&0&0&0&0\cr
           0&0&b(v)&0&0&0&c_{-}(v)&0&0\cr
           0&c_{+}(v)&0&b(v)&0&0&0&0&0\cr
           0&0&0&0&a(v)&0&0&0&0\cr
           0&0&0&0&0&b(v)&0&c_{-}(v)&0\cr
           0&0&c_{+}(v)&0&0&0&b(v)&0&0\cr
           0&0&0&0&0&c_{+}(v)&0&b(v)&0\cr
           0&0&0&0&0&0&0&0&w(v)} ,
\ee
where 
$a(v)=\sin(\gamma-v)$, $b(v)=\sin v$, $c_{\pm}(v)=e^{\pm{\i}v}\sin\gamma$
and $w(v)=\sin(\gamma+v)$.
Here $v$ is the spectral parameter and $\gamma$ (or $q=e^{-{\i}\gamma}$) 
is the crossing parameter.
This matrix is a trigonometric solution of the Yang-Baxter 
equation
\be
R^{12}(u-v)R^{13}(u)R^{23}(v)&=&R^{23}(v)R^{13}(u)R^{12}(u-v) .
\label{YBE}
\ee

On the other hand, the boundary $K$-matrices are given in terms of
\be
K(v,\xi)=\frac{1}{\sin\xi}\left(\begin{array}{ccc}
e^{-{\i}v}\sin(\xi+v)&0&0\\
0&e^{{\i}v}\sin(\xi-v)&0\\
0&0&e^{{\i}v}\sin(\xi-v)
\end{array}\right),\label{K} 
\ee
with \cite{Ruiz94}
\be
&&K^{-}(v)=K(v,\xi_-), \\
&&K^{+}(v) =q^{-1/2}K(-v+\gamma/2,\xi_+) M.
\ee
The crossing matrix $M$ is 
\be
M=\left(\begin{array}{ccc}1&0&0\\0&q^{2}&0\\0&0&-q^{2}
\end{array}\right) .
\ee
The $K$-matrices satisfy the boundary version of the Yang-Baxter 
equation \cite{Cherednik,Sklyanin}
\be
R_{12}(u-v)K_{1}^{-}(u)R_{21}(u+v)K_{2}^{-}(v)=
K_{2}^{-}(v)R_{12}(u+v)K_{1}^{-}(u)R_{21}(u-v). \label{sk1}
\ee

Following Sklyanin \cite{Sklyanin}, the open boundary condition transfer 
matrix is defined as \cite{FoKa:93,Ruiz94}
\be
\mbox{\boldmath$T$}(v)=\sum_{abcd} K_{ba}^{+}(v)
 U_{ac}(v)K^{-}_{cd}(v)U_{db}^{-1}(-v), \label{tr}
\ee
with $U_{ac}(v)$ the  monodromy matrix 
defined as the matrix product over the $R$'s,
\be
U_{ab(c)}^{\;\;(d)}(v)=R^{ad_{1}}_{b_{2}c_{1}}(v)
R^{b_{2}d_{2}}_{b_{3}c_{2}}(v)
R^{b_{3}d_{3}}_{b_{4}c_{3}}(v)...R^{b_{L}d_{L}}_{bc_{L}}(v).
\ee
Here the indices $c$ and $d$ in parentheses are in the quantum space 
${\bf C}^{3}\times {\bf C}^{3}\times ...\times{\bf C}^{3}$ with 
indices $a$ and $b$ in the horizontal auxiliary space ${\bf C}^{3}$ as usual.
The operator $U^{-1}(v)$ is the inverse of $U(v)$, with
\be
U_{ab(c)}^{-1(d)}(v)=\R^{b_{2}d_{1}}_{bc_{1}}(v)
\R^{b_{3}d_{2}}_{b_{2}c_{2}}(v)
\R^{b_{4}d_{3}}_{b_{3}c_{3}}(v)...\R^{ad_{L}}_{b_{L}c_{L}}(v),
\ee
where
\be
\R^{ab}_{cd}(v)=\frac{R^{ba}_{dc}(-v)}{\sin(\gamma+v)\sin(\gamma-v)}.
\ee
The elements of $\R$ will be denoted with a tilde, i.e., 
 $\;\tilde{a }\;,\tilde{b }\;,\tilde{ c}_\pm$ and $\tilde{w}$.

Given the above, the commutation relations 
\begin{equation}
[\mbox{\boldmath$T$}(v),\mbox{\boldmath$T$}(u)]=0  
\end{equation}
are fulfilled. The Hamiltonian $H$ (\ref{H}) follows from \cite{FoKa:93,Ruiz94}
\begin{equation}
{ \partial{\mbox{\boldmath$T$}}(v) \over \partial v} 
 \rule[-10pt]{0.2mm}{25pt}_{v=0}=
-\frac{1}{4}\;\sin\gamma\; H\;\; tr K^{+}(0)+tr \dot{K}^{+}(0).
\label{th}
\end{equation}

In the following we consider the quantum group $spl_q(2,1)$-invariant case only,
for which $H^\pm_s=0$. The transfer matrix (\ref{tr}) has been 
diagonalised for this and the more general case by means of the
algebraic Bethe Ansatz \cite{FoKa:93,Ruiz94}.
The eigenvalues are given by
\begin{equation}
\lambda(v)=\lambda_{\A}(v)+\lambda_{\D_{I}}(v)+\lambda_{\D_{II}}(v),
\label{eig}
\end{equation}
where
\be
\lambda_{\A}(v)&=&
\prod_{i=1}^{N}\frac{a(v_{i}-v)b(v_{i}+v)}
{b(v_{i}-v)a(v_{i}+v)}a^{L}(v)\a^{L}(-v)k_{\A}(v),\\
\lambda_{\D_{I}}(v)&=&k_{\D_{I}}(v)b^{L}(v)\b^{L}(-v)
\frac{b(2v)b(2v+\gamma)}{a(2v)w(2v+\gamma)}
\left(\frac{c_{+}(2v)}{a(2v)}-1\right)
\left(1-\frac{c_{-}(2v+\gamma)}{a(2v+\gamma)}\right)\no\\
&&\times\prod_{i=1}^{N}\frac{a(v-v_{i})\a(-v-v_{i}-\gamma)}
{b(v-v_{i})\b(-v-v_{i}-\gamma)}
\prod_{j=1}^{M}\frac{a(\nu_{j}-v)b(\nu_{j}+v+\gamma)}
{b(\nu_{j}-v)a(\nu_{j}+v+\gamma)},\\
\lambda_{\D_{II}}(v)&=& -k_{\D_{II}}(v)b^{L}(v)\b^{L}(-v)
\frac{b(2v)b(2v+\gamma)}{a(2v)w(2v+\gamma)}
\left(\frac{c_{+}(2v)}{a(2v)}-1\right)\no\\
&&\times\left(1-\frac{c_{-}(2v+\gamma)}{a(2v+\gamma)}\right)
\prod_{j=1}^{M}\frac{a(\nu_{j}-v)b(\nu_{j}+v+\gamma)}
{b(\nu_{j}-v)a(\nu_{j}+v+\gamma)} .
\ee
The related Bethe equations follow as
\be
&&\left(\frac{a(v_{k})\a(-v_{k})}{b(v_{k})\b(-v_{k})}\right)^{L}
\prod_{i\neq k}^{N}\frac{a(v_{i}-v_{k})b(v_{i}+v_{k})\b(v_{i}+v_{k}+\gamma)}
{a(v_{k}-v_{i})a(v_{i}+v_{k})\a(-v_{i}-v_{k}-\gamma)}\no\\
&&\hs{1}\times\prod_{j=1}^{M}\frac{a(\nu_{j}+v_{k}+\gamma)b(\nu_{j}-v_{k})}
{b(\nu_{j}+v_{k}+\gamma)a(\nu_{j}-v_{k})}=1,\hs{0.5} k=1,\ldots,N,
\label {BAE1}\\
&&\prod_{i=1}^{N}\frac{a(\nu_{l}-v_{i})\a(-\nu_{l}-v_{i}-\gamma)}
{b(\nu_{l}-v_{i})\b(-\nu_{l}-v_{i}-\gamma)}=1,\hs{1.2} l=1,\ldots,M.\label{BAE2}
\ee
The local boundary factors are
\be
k_{\A}(v)=k_{\D_{I}}(v)=k_{\D_{II}}(v)=1 . \label{k}
\ee

The three possible states ($\uparrow$,$\downarrow$,0) represent either
an electron (with spin up or down) or no electron (a hole). These are
described by the
numbers $N$ and $M$ of roots in the nested Bethe equations, 
where $M$ is the number of holes and $N-M$ is the number of down spins.
As for the periodic case \cite{FoKa:93a}, 
it follows that the magnetization  
$S^z=\frac{1}{2}(n_\uparrow-n_\downarrow)=\frac{1}{2}(L-2N+M)$ and the
number of electrons $Q=n_\uparrow+n_\downarrow=L-M$ are restricted to
$0 \le S^z \le Q/2 \le L/2$.
States in the half-filled band have one electron per site ($M=0$).

Our aim is to calculate the massless spin excitations of the states 
at or close to half-filling. 
This has been done for the periodic model in the isotropic ($\gamma \to 0$)
limit via the root density approach to the Bethe equations \cite{KaYa:91}. 
In Section 2 we establish an $su(2)$ structure for the open quantum group 
invariant model and derive the corresponding functional relations.
In Section 3 we derive the spin excitations by 
solving the functional relations for  
the finite-size corrections to the transfer matrix eigenspectra.
The bulk and surface free energies of the vertex model and related 
$t$--$J$ model are given in Section 4. 
A discussion and concluding remarks are given in Section 5.
 
\subsection{Functional relations }\setcounter{equation}{0}

We begin by considering the eigenvalue expression (\ref{eig}) from the
functional relation viewpoint.
To show the $su(2)$ structure we use semi-standard Young tableaux
as in the study of the six-vertex model with open boundaries \cite{Zhou:95}.

\subsubsection{The $su(2)$ structure}
Set ${\N}=2L$ and define
\be
{\sqbox{0}{-7}{\mbox{\scr $1$}}{}}^{k}  =
   \sin^{\N}(\gamma-k\gamma-v)\sin(2v\!-\!2\gamma\!+\!2k\gamma) 
  \prod_{i=1}^N{\sin(v-v_i+\gamma+k\gamma)\sin(v+v_i+k\gamma)\over
   \sin(v-v_i+k\gamma)\sin(v+v_i-\gamma+k\gamma)}, \hs{0.3} \label{box1} \\
{\sqbox{0}{-7}{\mbox{\scr $2$}}{}}^{k}\! =g_M(v\!+\!k\gamma)
    \sin^{\N}(v\!+\!k\gamma)\sin(2v\!+\!2k\gamma)
  \!\!\prod_{i=1}^N\!{\sin(v\!-\!v_i\!-\!\gamma\!+\!k\gamma)
      \sin(v\!+\!v_i\!+\!k\gamma\!-\!2\!\gamma)\over
       \sin(v\!-\!v_i\!+\!k\gamma)
      \sin(v\!+\!v_i\!-\!\gamma\!+\!k\gamma)}, \hs{0.3}\label{box2} \\
{\sqqbox{0}{-17}{1}{2}{\mbox{\scr $k$}}}=g_M(\!v\!+\!\gamma\!+\!k\gamma)
\sin^{\N}(\!\gamma\!\!-\!\!v\!\!-\!\!k\gamma\!)
 \sin^{\N}(v\!+\!k\gamma\!+\!\gamma)
  \sin(2v\!\!+\!\!2k\gamma\!\!-\!\!2\gamma)
  \sin(2v\!+\!2k\gamma\!+\!2\gamma), \hs{0.25}\label{box3} 
\ee\vskip 0.2cm

\noindent
where
\be
g_M(v)=\prod_{j=1}^M{\sin(v-\nu_j+\gamma)\sin(v+\nu_j-\gamma)\over
   \sin(v-\nu_j)\sin(v+\nu_j-2\gamma)}. \label{g}
\ee
The eigenvalue expression (\ref{eig}) now reads
\be
 \lambda(v)=\Bigl(\;\sqbox{0}{-7}{1}{}^{0}  +
      \; \sqbox{0}{-7}{2}{}^{0} \;\Bigr)
  {\sin^{-1}(2v-2\gamma)\over\sin^L(\gamma+v)\sin^L(\gamma-v)}
 -\sqqbox{0}{-17}{1}{2}{\mbox{\scr$\!-1$}}
  \;{\sin^{-1}(2v-2\gamma)\sin^{-1}(2v-4\gamma)
 \over \sin^{\N}(2\gamma-v)}.\label{lamT} 
\ee
To proceed further, introduce the auxiliary eigenvalues
\be
T^{(1)}_k  &=& {{\sqbox{0}{-7}{\mbox{\scr $1$}}{}}^{k}+
        \;       \sqbox{0}{-7}{\mbox{\scr $2$}}{}^{k}} \\
T^{(q+1)}_{0} & = & \sum  \begin{picture}(20,20)(-5,5)
    \put(0,0){${\underbrace
   {\sqbox{0}{3}{}{}
    \sqboxd{0}{3}{\!\!q\!+\!1}
    \sqbox{0}{3}{}{}}}^{\;0}$}
\end{picture} \label{Tq}
\ee\vskip 0.2cm

\noindent
where for each Young tableaux it is understood that there are relative shifts 
in the arguments:
\be
\sqbbox{0}{3}{\!\!\!\!v\!\!+\!\!q\!\gamma}{}
\sqbbox{0}{3}{...}{}
\sqbbox{0}{3}{...}{}
\sqbbox{0}{3}{\!\!\!v\!\!+\!\!\gamma}{}
\sqbbox{0}{3}{v}{\mbox{\scr$$}}^{\;0} \label{five-shifts}
\ee
\vskip -0.2cm\noindent
The zero superscript represents a shift in the right-most box. 
The number of terms in the sum is $(q+2)$, the dimension of 
the irreducible representations of $su(2)$. Namely they are given by
filling the numbers $1$ and $2$ in 
the $(q+1)$-box Young tableaux according to the rule that the numbers 
must not decrease moving to the right along the row. We thus get $q+2$ 
numbered Young tableaux. 
We can show that the auxiliary eigenvalues satisfy
\be
T^{(q)}_0T^{(1)}_q= T^{(q+1)}_0  + f_{q-1}T^{(q-1)}_0, 
\label{Func-T}  
\ee
or pictorially,
\be
&&\underbrace{\Atwon {180}{\otimes}{85}{q}} 
  \hs{0.85}{\Atwon {30}{}{5}{}}\no \\
&&\;=\;\underbrace{\Atwon {150}{\oplus}{55}{q-1}}\hs{0.85}
 \underbrace{\Atwon {210}{}{85}{q+1}} 
\ee
with
\be
f_k :=  \sqqbox{0}{-17}{1}{2}{\mbox{\scr $k$}} 
\ee

The functional relations (\ref{Func-T}) can be further used to show that
\be
T^{(q)}_0T^{(q)}_1&=&\prod_{k=0}^{q-1}f_k  + T^{(q+1)}_0T^{(q-1)}_1\;,
\label{T-p}
\ee
which coincides with (\ref{Func-T}) for $q=1$. It is also useful to introduce
\be     
y^q_0=\disp{T^{(q+1)}_0T^{(q-1)}_1/ \disp\prod_{k=0}^{q-1}f_k }\; .
\label{def-t}
\ee
with $y^0_0=0$. Then (\ref{T-p}) can in turn be used to show that
\be
y^q_0y^q_1=(1+y^{q+1}_0)(1+y^{q-1}_1)\;.\label{Func-t}
\ee
The relations (\ref{Func-T}) and (\ref{T-p}) are known as the $T$-system 
while (\ref{Func-t}) is the $y$-system 
\cite{KiRe:87,BaRe:89,KlPe:92,KNS:93}.

For half-filling ($M=0$) we have $S_z = \frac{1}{2} L - N$. Now  $g_M(v)=1$
and the function $f_k$ is independent of $M$.
It is quite clear from the $su(2)$ functional relations (\ref{Func-T}) that 
the term $f_0$ contributes only to the bulk and surface free energies rather 
than to the finite-size corrections of higher order. 
However, the relation 
$\partial T(0)/ \partial v \simeq \partial T^{(1)}(0)/ \partial v$
implies that the finite-size 
corrections to the eigenvalues of the transfer matrix $\lambda(v)$ are 
governed by the auxiliary eigenvalues $T^{(1)}_{0}$.
Thus the finite-size corrections to the Hamiltonian (1.1) follow from
the consideration of $T^{(1)}_{0}$. 
As in the study of the six-vertex model with open boundaries \cite{Zhou:95}, 
the surface effect on the finite-size corrections can
be calculated analytically from the $T$-system (\ref{Func-T}) and $y$-system
(\ref{Func-t}) functional relations. 

\subsubsection{Zeros and poles}\label{sec:zeros}

The functional relations have been shown to be very useful in calculating the 
finite-size corrections to the transfer matrices of exactly solved models
 \cite{KlPe:92,ZhPe:95,Zhou:95}. To solve the fusion hierarchy (\ref{Func-T})
and (\ref{Func-t}) we need to know the distribution of zeros and
poles of the auxiliary eigenvalues $T^{(q)}$ and $y^{(q)}$. 
We consider the model in the strip 
\be
-\gamma<{\rm Re}\;v<\gamma  \; . \label{strip}
\ee
Inside this strip the transfer matrix $T(v)$ in (\ref{tr}) is related to the
super-symmetric $t$--$J$ model (\ref{H}) via (\ref{th}). 
The clear advantage in working with the transfer matrix 
formulation is that it allows the application of powerful machinery 
from complex analysis.   
In this way we avoid the explicit manipulation of Bethe root densities, etc.
The largest, or groundstate, eigenvalue of $T^{(1)}$ is not expected to possess
zeros in the above strip. The zeros contributing to  
$T^{(q)}_0$ are of order $\N$ from the bulk. Those contributed by the boundary
are only of order $1$, which become unimportant in the limit $\N\to\infty$.
The bulk zero distribution is
\be
{\rm zero}[T^{(1)}(v)]&=&\emptyset,  \\
{\rm zero}[T^{(q)}(v)]&=&
    \bigcup_{k=0}^{q-2}\{-k\gamma\}^{\N} \hs{0.4}\mbox{for $q>1$}\,.
\ee
The zeros and poles of $y^{(q)}$ are determined by (\ref{def-t}), which gives 
\be
{\rm (I)}\;\;  q= 1\;:\hs{0.5}    
{\rm zero}[t^{(1)}(v)]&=& \{0\}^{\N},  \no\\
{\rm pole}[t^{(1)}(v)]&=&\{-\gamma\}^{\N}\{\gamma\}^{\N},   \\
{\rm (II)}\;\;  q\ge 2\;: \hs{0.3} 
{\rm zero}[t^{(q)}(v)]&=& \emptyset,\no\\
{\rm pole}[t^{(q)}(v)]&=&\{-q\gamma\}^{\N}\{\gamma\}^{\N},
\ee
for the bulk contribution only. Here the boundary contribution to the
zeros and poles, of order greatly less than $L$, are not listed and 
contribute less than those of
order $\N$ when the system size $L$ becomes large. 
Only these zeros or poles of order $L$ are especially important
in the thermodynamic limit $L\to\infty$ \cite{Zhou:95}.  

\subsubsection{The functional relations for finite-size corrections}
\label{finte}
The finite-size corrections to the eigenvalues $T^{(1)}$ can be 
obtained by solving the functional relations (\ref{Func-T}) and 
(\ref{Func-t}) in the strip (\ref{strip}). 
Denote the finite-size corrections of $T^{(1)}$
by $T^{(1)}_{\mbox{\tiny finite}}(v)$ and write
\be
T^{(1)}(v)=T^{(1)}_{\mbox{\tiny finite}}(v)
  T^{(1)}_{\mbox{\tiny free}}(v) .\label{def-finite}
\ee
The bulk and the surface free energy contributions together satisfy
\be
T^{(1)}_{\mbox{\tiny free}}(v)T^{(1)}_{\mbox{\tiny free}}(v+\gamma)
=f_0 . \label{free}
\ee
Inserting (\ref{def-finite}) into (\ref{T-p}) or (\ref{Func-T}) we find that
\be
T^{(1)}_{\mbox{\tiny finite}}(v)T^{(1)}_{\mbox{\tiny finite}}(v+\gamma)
=1+y^{(1)}(v) .
\label{finite}\ee
The finite-size  corrections to $T^{(1)}(v)$ are thus represented by
the $y$-system component $y^{(1)}(v)$. In the following we give an analytical
treatment  of (\ref{finite}) and (\ref{Func-t}). We will see that 
the finite-size corrections in the scaling limit are dependent only on
the braid asymptotics and the bulk behavior of the functional relations.

The Bethe equations (\ref{BAE1})-(\ref{BAE2}) render $T^{(1)}(v)$ analytic. 
Since all functions involved in the eigenvalues are $\pi$-periodic, the 
analyticity domains for $T^{(1)}(v)$ are not  unique. It is thus useful to 
introduce functions of a real variable by restricting the
eigenvalue functions to certain lines in the complex plane,
\be
\hs{1.2}{\cb}(x)&:=&T^{(1)}_{\mbox{\tiny finite}}
   \({{\i}\over \pi}x\gamma +{1\over 2}\gamma\) ,\label{nf1} \\
{\al}^{(q)}(x)&:=&y^{(q)}\({{\i}\over \pi}x\gamma +{1-q\over 2}\gamma\) 
, \label{nf2} \\ 
{\U}^{(q)}(x)&:=&1+{\it \al}^{(q)}(x) .\label{nf3}
\ee
For the groundstate the functions $\U^{(1)}(x)$ and $\cb(x)$ 
are {\em analytic, non-zero} (for those of order $\cal L$)
in $-\pi<{\rm Im}\;x<\pi$ and  possess {\em constant}
asymptotics for ${\rm Re}\;x\to \pm \infty$ (the ANZC property), 
which can be seen directly from the eigenvalues. 

Eqn (\ref{finite}) can be solved using the new functions and applying 
Fourier transforms,
\be
{F_T}(k)&=&{1\over 2\pi}\int_{-\infty}^{\infty} dx\;[\ln\cb(x)]'
    \;e^{-{\i} kx} ,\no\\
\h [\ln\cb(x)]'&=&\int_{-\infty}^{\infty} dk\;{F_T}(k) \;e^{{\i} kx}, \\
{\U^{(q)}}(k)&=&{1\over 2\pi}\int_{-\infty}^{\infty} dx\;[\ln F_A^{(q)}(x)]'
    \;e^{-{\i} kx} ,\no\\
\h [\ln F_A^{(q)}(x)]'&=&\int_{-\infty}^{\infty} 
  dk\;{\U^{(q)}}(k) \;e^{{\i} kx} .
\ee
We obtain
\be
\ln\cb=k*\ln \U^{(1)} , \label{b}
\ee
where the kernel $k(x)$ is 
\be
k(x):= {1\over 2\pi\cosh(2x)} \,.
\ee
Here the convolution $f*g$ of two functions $f$ and $g$ is defined by
\be
(f*g):=\int^\infty_{-\infty}f(x-y)g(y)\;dy= 
       \int^\infty_{-\infty}g(x-y)f(y)\;dy\; . 
\label{convolution}\ee 
There is an integration constant $C$ in (\ref{b}) which we drop
because it does not contribute to the $1/L$ corrections.  
In case of the low-lying excitations we have to take care of zeros 
in the analyticity strips so that the simple ANZC properties hold. The result 
(\ref{b}) is still correct if we change the integration path ${\cal L}$ so that 
$\cb(x)$ is ANZC and the Cauchy theorem can be applied as 
discussed elsewhere \cite{KlPe:92,Zhou:95}. 

The function $\U^{(1)}$ is determined by the $y$-system (\ref{Func-t}).
According to Section~\ref{sec:zeros} the analyticity strip (\ref{strip})
for $y^{(1)}(v)$ contains a zero of order $\N$ at $u=0$ and 
a pole  of order $\N$ at $u=\pm\gamma$.  
All other functions $y^{(q)}$ are  analytic and non-zero in
their analyticity strips $-\gamma<u<\gamma$. 
Taking care of these properties, applying Fourier transforms to the
logarithmic derivative of the equations (\ref{Func-t}) with the new
functions (\ref{nf1})-(\ref{nf3})
and then integrating the equations back we obtain the nonlinear integral 
equations
\be
\ln\al^{(q)}=\ln\e^{(q)}+k*\ln\U^{(q-1)}+k*\ln\U^{(q+1)}+D^{(q)} , \label{a}
\ee
where
\be
\e^{(q)}(x):=\left\{ \begin{array}{ll}        \label{e}
1 \; , & q\not=1 \\
\tanh^{\mbox{\scr$\cal L$}}(\half x)\; , & q=1 .
\end{array}\right.
\ee
Here $D^{(q)}$ are integration constants. For the same reason we have to take 
care of the ANZC property in the analyticity strips in (\ref{a}). 

\subsubsection{The functional relations in the limit $L\to\infty$}

The finite-size corrections can be extracted from the
nonlinear integral equations (\ref{a}) and (\ref{b}). 
The system size $L$ enters the nonlinear equations (\ref{a}) through
(\ref{e}). The function $\e^{(1)}$ has three asymptotic regimes with 
transitions 
in scaling regimes when $x$ is of the order of $-\ln L$ or $\ln L$.
We suppose that $\al^{(q)}$ and $\U^{(q)}$ scale similarly. Thus in the 
following 
scaling limits,
\be
&&e^{(q)}_{\pm}(x):=\lim_{{L}\to \infty}\e^{(q)}\(\pm(x+\ln {\N})\)  ,   \no \\
&&a^{(q)}_{\pm}(x):=\lim_{{L}\to \infty}\al^{(q)}\(\pm(x+\ln {\N})\)  ,\\
&&A^{(q)}_{\pm}(x):=\lim_{{L}\to \infty}\U^{(q)}\(\pm(x+\ln {\N})\)
=1+a^{(q)}_\pm(x) ,\no  
\label{scaling}
\ee
eqn (\ref{a}) takes the form
\be
\la^{(q)}=\Le^{(q)}+k*\lA^{(q-1)}+k*\lA^{(q+1)}+D^{(q)}  ,\label{a-L}
\ee 
where we use the abbreviations
\be
&& \la^{(q)}(x):=\ln a^{(q)}(x) \; ,\h \lA^{(q)}(x):=\ln A^{(q)}(x)  , \no\\
&& \Le^{(q)}(x):=\left\{\begin{array}{ll}
                    0\; ,      &q\not=1 \,,\\
                    -2e^{-x}\;,&q\not=1 \,,  \end{array}
             \right. \label{Le}
\ee
and suppress the $\pm$ subscripts. 
The transfer matrix $\cb(x)$ in the $L\to\infty$ limit now becomes
\be
\ln\cb(x)&=&(k*\ln\U^{(1)})(x) \no \\
&=&{1\over 2\pi}\int^\infty_{-\ln {\cal N}}\({\ln\U^{(1)}(y+\ln {\N})\over
 \cosh(x-y-\ln {\N})}+{\ln\U^{(1)}(-y-\ln {\N})\over
 \cosh(x+y+\ln {\N})}\)\;dy  +{ o\!}\({1\over {\N}}\) \no\\
&=&{e^x\over {\N}\pi}\int^\infty_{-\infty} e^{-y}\lA^{(1)}_+(y)\;dy
+{e^{-x}\over {\N}\pi}\int^\infty_{-\infty} e^{-y}\lA^{(1)}_-(y)\;dy
+{ o\!}\({1\over {\N}}\)  \no \\
&=&{2\cosh x\over {\N}\pi}\int^\infty_{-\infty} e^{-y}
\lA^{(1)}(y)\;dy+{ o\!}\({1\over {\N}}\)  .\label{b-N}
\ee
The above equation converges and can actually  be
evaluated explicitly with the help of the dilogarithmic function
\be
L(x)=-\int_0^x dy\;{\ln (1-y)\over y} +\half \ln x\ln (1-x) .
\ee
Multiplying the derivative of (\ref{a-L}) with $\lA^q$, and (\ref{a-L})
itself with $(\lA^q)'$, taking the difference, summing over $q$
and using (\ref{Le}), we are able to obtain
\be
2\int_{-\infty}^\infty e^{-y}\lA^{(1)}(y)\;dy=
-\sum_{q}^{}L\({1\over A^{(q)}}\)\rule[-15pt]{0.2mm}{35pt}_{-\infty}^\infty+
\half\sum_{q}^{}D^{(q)}\lA^{(q)}\;\rule[-15pt]{0.2mm}{35pt}_{-\infty}^\infty,
 \label{integral}
\ee
where the constants $D^{(q)}$ are given in terms of the $x\to\infty$ 
asymptotics by
\be
D^{(q)}=\la^{(q)}-\half \lA^{(q-1)}-\half \lA^{(q+1)} . \label{D}
\ee
The result (\ref{integral}) shows that 
the finite-size corrections in the scaling limit depend only on
the braid asymptotics and the bulk behaviour of the functional relations.

\subsubsection{Asymptotics and bulk behavior}\label{Asmpsec}

The nonlinear integral equations (\ref{a-L}) can be easily
solved in the limit $x\to\pm\infty$ with  
\be
\gamma={\pi\over h}\h h=3,4,\cdots\;. \label{gamma}
\ee
In many cases different $h$ correspond to different models. The equation
(\ref{integral}) shows that these asymptotic solutions are enough 
to obtain the finite-size corrections of the transfer matrix $T^{(1)}(v)$.

It is obvious to see that the $x\to\infty$ asymptotics corresponds to the 
braid limit  $u\to\pm{\i}\infty$. In this limit (\ref{Func-t}) reduces to
\be
(y^{(q)}_\infty)^2=(1+y^{(q-1)}_\infty)(1+y^{(q+1)}_\infty)  .\label{rec}
\ee
This equation in turn means
\be
2\la^{(q)}=\lA^{(q-1)}+ \lA^{(q+1)}+D^{(q)},
\ee
in terms of the functions $a^q$. The constants 
$D^{q}$ can be either zero or non-zero as the 
different branches can be taken for the logarithmic functions 
in the nonlinear integral equations.  

To solve for $y^{(q)}_\infty$ we write $y^{(1)}_\infty$ as
\be
y^{(1)}_\infty={\sin 3\theta \over \sin \theta}, \label{theta}
\ee
where the parameter $\theta$ is to be determined. The recursion relation 
(\ref{rec}) implies 
\be
y^{(q)}_\infty&=&{\sin q\theta\, \sin (q+2){\theta}
/ \sin^2\theta}, \no\\
y^{(q)}_\infty+1&=&{\sin^2 (q+1)\theta\,/ \sin^2\theta},
\ee
for all $q=1,2\cdots$. This solution has to be consistent with the
braid limit of $T^{(q)}(v)$. To fix the constant parameter 
$\theta$ let us consider the groundstate, for which $N=\half (L+M)$ and
\be
\lim_{{\rm Im}v\to\pm\infty}T^{(1)}(v)/\phi(v)=2\cos\({\pi\over h}\)\;.
\ee 
Recalling (\ref{theta}) and using the relations 
\be
y^{(1)}_\infty&=&\lim_{{\rm Im}u\to\pm\infty}{y^{(2)}_0/ f_0} \no \\
&=&\lim_{{\rm Im}u\to\pm\infty}{T^{(1)}_0T^{(1)}_1/ f_0}-1 \no \\
&=& 4\cos^2({\pi/ h})-1
\ee
we have 
\be
\theta=\gamma={\pi\over h}.
\ee
Moreover, the special values of $\theta$ lead to the closure condition
\be
y^{(h-2)}_\infty=0 \;. \label{clo}
\ee
For the sector $S^z=\half (L+M)-N$ we have to modify $\theta$ to be
\be
\theta=m\gamma={m\pi\over h}\;
\ee
where $m=2 S^z +1=1,3,\cdots\le h-1$. 

In the limit $x\to-\infty$, $y^{(q)}$ can be considered as the bulk
behaviour in large $L$. According to Section~\ref{sec:zeros} the 
analyticity strip for $y^{(1)}(v)$ contains a zero of order $N$ at 
$v=0$ and poles of order $N$ at $v=-\pm\gamma$.
All other functions $y^{(q)}$ are analytic and non-zero in their 
analyticity strips in $-\gamma< v<\gamma$. 
For large $N$ we find that the leading bulk behaviour to $y^{(q)}$ is
\be
y^{(q)}_{\rm bulk}(v)=\left\{ \begin{array}{ll}
 \mbox{constant ,}                            & q\not=1 \; ,\\
 \mbox{constant} \left[\tan({1\over 2}h v)\right]^{\cal L}\; , & q=1 \; .
\end{array} \right. \label{bulk-t}\ee
The constants are fixed by the functional equations (\ref{Func-t}) 
and can be calculated similarly to the asymptotics of $y^{(q)}$. 
As for the ABF model \cite{KlPe:92}, it is easy to see that the limit 
\be
\lim_{x\to-\infty}\lim_{N\to\infty}y^{(1)}\sim \lim_{x\to-\infty} 
\exp{(-2e^{-x})} = 0\;.
\ee 
Therefore the functional equations (\ref{Func-t}) are modified 
and we find the constants for $2\le q\le h-3$, with 
\be
y^{(q)}_{\rm bulk}&=&{\sin (q-1)\tau\, \sin (q+1){\tau}
   / \sin^2\tau},\no\\
y^{(q)}_{\rm bulk}+1&=&{\sin^2 q \tau/ \sin^2\tau} , \label{bulk}
\ee
where
\be
&& \tau={m^{''}\pi\over h-1}    ,
\ee
which is consistent with the closure condition (\ref{clo}). 
 
The transfer matrix eigenspectra has only one ``quantum number" $S^z$.
There should be one free parameter between $m^{'}$ and $m^{''}$. 
For the largest (groundstate) eigenvalue 
the appropriate choices are $m=m^{''}=1$. 
The  open boundary $t$--$J$ system under consideration possesses 
$spl_q(2,1)$ invariance. In the case of a fixed $M$ the Bethe 
equations and transfer matrix eigenvalues $T^{(1)}$ are similar to
those of the six-vertex model with open $su_q(2)$-invariant boundaries. Thus
we suppose that $m^{''}=1$ as in the study of the $XXZ$-chain 
\cite{HQB:87,SaBa:89} 
and the related six-vertex model \cite{Zhou:95}. 
The low-lying excited states are given by $m>1$.

The solution $y^{(1)}_{\rm bulk}(v)$ is given by
\be
y^{(1)}_{\rm bulk}(v)y^{(1)}_{\rm bulk}(v+\gamma)&=&(1+y^{(2)}_{\rm bulk})  
\no \\
&=& 4 \cos^2 \tau .
\ee
Thus we find lastly that
\be
y^{(1)}_{\rm bulk}(v)=\pm 2\cos\tau\left[\tan({1\over 2}h v)\right]^{\N} .
\ee

\subsection{Finite-size corrections}\setcounter{equation}{0}

The finite-size corrections are only dependent on the braid and bulk
limits. In these limits the functional relations are truncated
and the summation in (\ref{integral}) can be replaced with 
\be
2\int_{-\infty}^\infty e^{-y}\lA^{(1)}(y)\;dy=
-\sum_{q=1}^{h-3}L\({1\over A^{(q)}}\)
 \rule[-15pt]{0.2mm}{35pt}^{\infty}_{-\infty\;\; .}+
 \half\sum_{q=1}^{h-3}D^{(q)}
 \lA^{(q)}\;\rule[-15pt]{0.2mm}{35pt}_{-\infty}^\infty .
 \label{integra2}\ee
Recall that the constants $D^{(q)}$ are dependent on the 
branches of the dilogarithmic functions in 
the nonlinear integral equations. The appropriate choice yields the 
correct finite-size corrections. Simply taking $D^{(q)}=0$ 
is consistent with the asymptotics solutions given in 
Section~\ref{Asmpsec}. To take nonzero
$D^{(q)}$ we need to single out the appropriate branches from 
the asymptotic solutions of the equations, as has been shown 
for ABF models \cite{KlPe:92}.  

A relevant dilogarithm identity has been established by
Kirillov \cite{Kirillov:93}. Consider the functions
\be
 y^{(q)}\!(j,r):={\sin(q+2)\varphi\;\sin(q\varphi)\over
          \sin^2(\varphi)}\; ,
 \h 1\le b\le n-1,\h 1\le q\le r, \label{su(n)2}
\ee 
with
\be
\varphi={(1+j)\pi\over 2+r }\h 0\le j\le r\;.
\ee 
It is obvious that they represent the solutions of the asymptotic 
equations (\ref{a-L}) with $r=h-2$ or the solutions of (\ref{bulk}) for
the bulk behaviour with $r=h-3$.
Then we have the dilogarithm identity 
\be s(j,r)&:=&
\sum_{q=1}^{r}L({1\over 1+y^{(q)}\!(j,r)})
      \label{dilogarithm-s}      \no\\ 
&=&{\pi^2 \over 6}\left[{3r\over 2+r}-{6j(j+2)\over 2+r} +\;6j\right] .   
\label{s}
\ee

Now in terms of the dilogarithm function the finite-size corrections 
(\ref{b-N}) are expressed as 
\be
\ln\cb(x)={\cosh x\over \N\pi}\left[\;\;\sum_{q=1}^{h-3}
  L\({1\over A^{(q)}}\) \rule[-15pt]{0.2mm}{35pt}_{-\infty}^\infty+
 \half\sum_{q=1}^{h-3}D^{(q)}
 \lA^{(q)}\;\rule[-15pt]{0.2mm}{35pt}_{-\infty}^\infty\;\;\right]
 +{ o\!}\({1\over {\N}}\)\label{bb} .
\ee
Note that the nonlinear
integral equations (\ref{a-L}) including the closure condition
(\ref{clo}) and their solutions presented in 
Section~\ref{Asmpsec} are the same as those of the ABF models \cite{KlPe:92}. 
Therefore we can calculate the finite-size corrections in the same way. 
Similarly to \cite{KlPe:92}, with $D^{(q)} \ne 0$, it can thus be shown  
that in terms of the functions $s(j,r)$ the finite-size corrections
(\ref{bb}) for the quantum-invariant open boundary system can be written   
\be
\ln\cb(x) &=&{\pi\cosh x\over 6\N} \left[
                    s(0,h-3)+s(0,1)-s(m-1,h-2) \right. \no\\
&&\hs{2}\left. -{6(1-m)(2-m)}\right]+{ o\!}\({1\over {\N}}\) .
\label{excitation1}
\ee 
Inserting (\ref{s}) into this equation we have the
finite-size corrections in the more recognizable form \cite{Cardy:86}
\be
\ln\cb(x)={\pi\over 6\N}\(c-24\Delta_{m}\)\cosh x
 +{ o\!}\({1\over \N}\), \label{b-cd}
\ee
where the central charge and conformal weights are given by 
\be
         c&=&1-{6\over h(h-1)},  \label{c} \cr 
\Delta_{m}&=&{\left[ h-(h-1)m \right]^2-1\over 4h(h-1)},  \label{delta}
\ee
with $m=1,3,\cdots\le h-1$. 

\subsection{Free energies}\label{fe}

Consider now the bulk and surface free energies. Let 
\be
\lambda_{\mbox{\tiny free}}(v)=T_b(v)T_s(v)=
\displaystyle{{T^{(1)}_{\mbox{\tiny free}}((v)\over\sin(2v-2\gamma)
\sin^L(\gamma+v)\sin^L(\gamma-v)}},
\ee
where $T_b$ and $T_s$ are the bulk and surface contributions.
They are determined by (\ref{free}). Factors of order $L$ in $f_0$ 
contribute to $T_b$ and otherwise to $T_s$. 
From (\ref{lamT}) and (\ref{free}) we should have
\be
T_b(v)T_b(v+\gamma)=
  {\sin^{L}(\gamma+v)\sin^{L}(\gamma-v)
    \over \sin^{2L}(\gamma)}{\sin^{2L}(\gamma)\over
 \sin^{L}[\gamma+(\gamma+v)]\sin^{L}[\gamma-(\gamma+v)]} \label{inv-b34}
\ee
for the bulk and 
\be
T_s(v)T_s(v+\gamma)=
  {\sin(2v+2\gamma)\sin({2\gamma-2v})\over\sin^2(2\gamma)}{
  \sin^2(2\gamma)\over\sin[\gamma+(\gamma-2v)]\sin[\gamma-(\gamma-2v)]} 
\label{uni}
\ee
for the surface. Solving these equations we find 
\be
\log T_b(v)&=&  {2L}   
 \int_{-\infty}^{\infty} dk {\cosh( 2 \gamma k-\pi k)\sinh( v k)
\sinh(\gamma k)\cosh( v k)\over k\cosh(\gamma k)\sinh(\pi k)}, 
\label{Tb}\\
\log T_s(v)&=&\int_{-\infty}^{\infty} dk{\cosh(4\gamma k-\pi k)\sinh(2 vk)
\sinh(2\gamma k-2 v k)\over 
   k\cosh(2\gamma k)\sinh(\pi k)} \no\\
&&-\int_{-\infty}^{\infty} dk{\cosh(2\gamma k-\pi k)\sinh(2vk)
  \sinh(\gamma k-2 v k)\over 
   k\cosh(\gamma k)\sinh(\pi k)}. \label{Ts}
\ee

\subsubsection{$t$--$J$ model}

The eigenvalues $E$ of the quantum invariant $t$--$J$ Hamiltonian (\ref{H}) 
follow via the relation (\ref{th}), with  
\be
E_m=-{4\over \sin\gamma }{\partial{\lambda}(v)\over\partial v}
 \rule[-10pt]{0.2mm}{25pt}_{v=0} \, .
\ee
From the above, we have that the groundstate energy is given by
$E_0=2L\;e_b+2 e_s+e_s^+ +e_s^-$, where
\be
e_b&=& -{2\over L\; \sin\gamma }{\partial{\log T_b}(v)
  \over\partial v}\rule[-10pt]{0.2mm}{25pt}_{v=0} \no \\
&=&-{4\over \sin\gamma }\int_{-\infty}^{\infty} 
 dk {\cosh( 2 \gamma k-\pi k)\sinh( \gamma k)
    \over \cosh(\gamma k)\sinh(\pi k)}, \label{Eb} 
\ee
and
\be
e_s&=& -{2\over \sin\gamma }{\partial{\log T_s}(v)
  \over\partial v}\rule[-10pt]{0.2mm}{25pt}_{v=0} \no\\
&=&-{4\over \sin\gamma }\int_{-\infty}^{\infty} 
 dk {\cosh( 4 \gamma k-\pi k)\sinh(2 \gamma k)
    \over \cosh(2\gamma k)\sinh(\pi k)}  \no \\
&&+{4\over \sin\gamma }\int_{-\infty}^{\infty} 
 dk {\cosh( 2 \gamma k-\pi k)\sinh( \gamma k)
    \over \cosh(\gamma k)\sinh(\pi k)}. \label{Es}
\ee
Here the local surface energies $e_s^\pm=0$ for the 
quantum invariant case. 
At half-filling, the results for $e_b$ and $e_s$ are in agreement
with those of the XXZ chain \cite{HQB:87}, as expected. 
The spin excitations are given by
\be
E_m&=&E_0 -{4\over \sin\gamma }
  {\partial{\log T_{\mbox{\tiny finite}}}(v)
  \over\partial v}\rule[-10pt]{0.2mm}{25pt}_{v=0}\no\\
&=&E_0 -{v_s\pi\over 24 L }(c-24\Delta_m)+{o}
   \bigl({1\over L}\bigr),
\ee
where the sound velocity is 
\be
v_s=\cases{\displaystyle{\pi\over 3\gamma\sin\gamma} 
    &$\gamma\not=0$ ,\cr
  \displaystyle{\pi\over 3 } &$\gamma=0$  .\cr}
\ee

\subsection{Conclusion and discussion}\setcounter{equation}{0}

We have exploited the $su(2)$ structure of the transfer matrix 
functional relations to calculate the massless spin excitations 
of the integrable quantum group invariant $t-J$ model (\ref{H}) 
at close to half-filling. We took the special value $\gamma=\pi/h$ of the 
anisotropy parameter in order to close the functional relations.
Results for the isotropic model are recovered in the limit $h\to\infty$.
Explicity, in this limit the scaling dimensions of the spin excitations, 
or spinons, follow from (\ref{delta}) as
\be
X_{S_z} = 2 \Delta_{S_z} = 2 (S^z)^2, \label{spinons}
\ee 
where $S^z=0,1,\cdots\ $.  

\subsubsection{General parameter $0\le\gamma\le \pi/2$}

The excitation spectrum can be obtained for
general anisotropy parameter $\gamma$ via a related analytic
nonlinear integral equation approach 
\cite{KBP:91,WBN:92,KWZ:93,Zhou:95a,ZhBa:96}.
For the quantum invariant model at or close to half-filling we expect 
\be
\log\lambda(v)=\log T_b +\log T_s +{\pi\over 6\N}
 \(c-24\Delta_{m}\)\sin({\pi v\over \gamma})
  +{ o\!}\({1\over \N}\),
\ee
with the bulk and surface free energies given by (\ref{Tb})
and (\ref{Ts}). The central charge and conformal dimensions are
\be
c&=&1-{6\gamma^2\over \pi(\pi-\gamma)}, \label{c1} \\
\Delta&=&{[\gamma m-(m-1)\pi]^2-\gamma^2\over 4\pi(\pi-\gamma)}, 
\label{delta1}
\ee
where $m=2S^z+1=1,3,\cdots$. It follows that the
finite-size corrections to the anisotropic $t-J$ model are 
\be
E_m&=&2L e_b+2e_s-{\pi v_s\over 6\N}\(c-24\Delta_{m}\)
  +{ o\!}\({1\over \N}\) \label{Em}
\ee
where the bulk and surface terms are given by (\ref{Eb}) and (\ref{Es}).

\subsubsection{Local surface energies}

The above results are for vanishing boundary fields $H^\pm_s$.
The quantum group $spl_q(2,1)$ invariance is broken for $H^\pm_s\not=0$. 
However, the model remains integrable if \cite{Ruiz94}
\be
H_s^-&=&{\i}\sin\gamma \;(\cot\xi_- -1)(S^{z}_{1}-n^{h}_{1}/2) \\
H_s^+&=&-{\i}\sin\gamma \;(\cot\xi_+ -1)(S^{z}_{L}-n^{h}_{L}/2)
\ee
where $H^\pm_s=0$ is recovered in the limit $\xi_\pm\to\infty$.
For finite $\xi_\pm$ the eigenvalues and Bethe 
equations are given by (\ref{eig})-(\ref{BAE2}) with \cite{Ruiz94}
\be
k_{\A}(v)&=&
\frac{q\sin(\xi_{+}-v)\sin(\xi_{-}+v)}{\sin\xi_{+}\sin\xi_{-}},\\
\nonumber\\
k_{\D_{I}}(v)&=&\frac{\sin(\xi_{+}+v-\gamma)
\sin(\xi_{-}-v+\gamma)}{\sin\xi_{+}\sin\xi_{-}},\\
\nonumber\\
k_{\D_{II}}(v)&=&
\frac{\sin(\xi_{-}-v+\gamma)
\sin(\xi_{+}+v-\gamma)}{\sin\xi_{+}\sin\xi_{-}}.
\end{eqnarray}
At or close to half-filling the free energy can be calculated in a similar
manner to that presented here for the quantum invariant case. 
As seen in Section~\ref{finte}, the total groundstate energy of the $t$-$J$
model satisfies (\ref{free}).  The local surface free energy, 
on the other hand, satisfies
\be
T_s^\pm(v)T_s^\pm(v+\gamma)=
  {q\sin(\xi_\pm+v)\sin(\xi_\pm-v)
    \over \sin(\xi_\pm)\sin(\xi_\pm)}. \label{inv-loc}
\ee
Here we omit the factor $q$ in the following as it
shifts $\log T_s^\pm(v)$ with no contribution to the surface energy of 
the quantum chain. We find
\be
\log T_s^\pm(v)&=& 
 \int_{-\infty}^{\infty} dk 
  {e^{ k(\gamma-\xi_\pm)}\cosh( 2 \xi_\pm k-\pi k)
   \sinh( v k)
\sinh(\xi_\pm k- v k)\over k\cosh(\gamma k)\sinh(\pi k)} 
\ee
Hence the local surface energy of the quantum chain is given by
\be
e_s^\pm&=&-{4\over tr K^{+}(0)\sin\gamma}{\partial{\log T_s^\pm}
  (v)\over\partial v}
 \rule[-10pt]{0.2mm}{25pt}_{v=0}  \no\\
&=&-\displaystyle{{4\over tr K^{+}(0)\sin\gamma}
 \int_{-\infty}^{\infty} dk 
{e^{ k(\gamma-\xi_\pm)}\cosh(2\xi_\pm k-\pi k)
   \sinh(\xi_\pm k)\over  \cosh(\gamma k)\sinh(\pi k) }}\,.
\ee

\subsubsection{Hole excitations}

We have considered spin excitations, the so-called spinons,
at or close to the half-filled band. The spinon excitations
are given by the dimensions (\ref{delta}) and (\ref{delta1}),
with (\ref{spinons}) for the isotropic model, where 
$S^z = \half (L-2N+M)$. One also needs to 
consider the holon part of the spectrum. 
According to the results obtained for the excitation spectra of
the periodic $t$--$J$ model, the holons should 
contribute another independent conformal theory \cite{KaYa:91}.
In contrast to the spinon part, the calculation of the free 
energies from $\lambda(v)$ is no longer as per Section~\ref{fe}.
Let us write 
$\lambda(v)=\lambda_{\mbox{\tiny free}}(v)\lambda_{\mbox{\tiny
finite1}}(v) \lambda_{\mbox{\tiny finite2}}(v)$, then $\lambda_{\mbox{\tiny
finite1}}(v)=T^{(1)}_{\mbox{\tiny finite}}(v)$. Define 
$\lambda_{\mbox{\tiny free,finite2}}(v)=\lambda_{\mbox{\tiny free}}(v)
\lambda_{\mbox{\tiny finite2}}(v)$. Both the bulk and surface free energies 
should follow from
\be
\lambda_{\mbox{\tiny free,finite2}}(v)\lambda_{\mbox{\tiny free,finite2}}
(v+\gamma)=d(v)g_M(v+\gamma),   \label{la}
\ee
which is clearly dependent on $M$, where
\be 
d(v)={\sin^{L}(\gamma+v)\sin^{L}(\gamma-v)\over \sin^{L}(2\gamma+v)\sin^{L}(-v)}
 {\sin(2\gamma+2v)\over \sin(2v)}.
\ee
This shows that the bulk and surface 
free energies follow from two parts, namely $g_M(v)$ and $d(u)$.
The energies $e_b$ and $e_s$ given in (\ref{Eb}) and (\ref{Es}) are
the contribution from $d(v)$. 
The above relation also encodes the
finite-size corrections to the eigenvalues, and thus the holon part
of the spectra, through the contribution $\lambda_{\mbox{\tiny finite2}}(v)$. 
It is obvious that we need to analyse
the Bethe equation (\ref{BAE1})-(\ref{BAE2}) to solve the inversion 
relation (\ref{la}). It is possible that we will need to use a very different
method to obtain the holon excitations.
Whether or not the central charge
associated with the holon part of the spectrum is also less than one 
for the anisotropic quantum invariant $t$--$J$ model remains to be explored.  

\vskip 5mm
\noindent
{\em Note Added.} After completing this work we received a preprint
by Asakawa and Suzuki in which the
finite-size corrections are calculated for the open isotropic $t$--$J$ 
model via the root density method \cite{AsSu:96}. 
They find the central charge $c=1$ for both spinons and holons in agreement
with the periodic case \cite{KaYa:91}. The conformal weights of the
spinons agree with our result (\ref{spinons}). In addition to treating
the corresponding vertex model our results 
for the spinon conformal spectra generalise those of Asakawa and Suzuki 
to the anisotropic quantum invariant $t$--$J$ model.  
 
\subsection*{Acknowledgements} This research has been supported 
by the Australian Research Council and the Natural Science
Foundation of China.

\end{document}